\documentclass[10pt,preprint]{aastex}
\begin{document}

\title{THERMAL-IR DETECTION OF OPTICAL OUTFLOW SOURCES IN OMC1 SOUTH\altaffilmark{1}}

\author{Nathan Smith\altaffilmark{2,3}, John Bally\altaffilmark{3},
Ralph Y.\ Shuping\altaffilmark{4}, Mark Morris\altaffilmark{4}, and
Thomas L.\ Hayward\altaffilmark{5}}

\altaffiltext{1}{Based on observations obtained at the Gemini
Observatory, which is operated by the Association of Universities for
Research in Astronomy, Inc., under a cooperative agreement with the
NSF on behalf of the Gemini partnership: the National Science
Foundation (US), the Particle Physics and Astronomy Research Council
(UK), the National Research Council (Canada), CONICYT (Chile), the
Australian Research Council (Australia), CNPq (Brazil), and CONICET
(Argentina).}

\altaffiltext{2}{Hubble Fellow; nathans@casa.colorado.edu}

\altaffiltext{3}{Center for Astrophysics and Space Astronomy, University of
Colorado, 389 UCB, Boulder, CO 80309}

\altaffiltext{4}{Division of Astronomy and Astrophysics, University of
California at Los Angeles, Los Angeles, CA 90095}

\altaffiltext{5}{Gemini Observatory, Association of Universities for
Research in Astronomy, Inc., Casilla 603, La Serena, Chile}

\begin{abstract}

We present the first thermal-infrared imaging photometry for several
embedded sources in the OMC1 South cloud core in the Orion nebula, and
propose that some of these drive optical Herbig-Haro jets emerging
from the region.  Thermal-infrared images at 8.8 and 11.7~$\micron$
obtained at Gemini South show a handful of sources in OMC1-S with no
visual-wavelength counterparts, although a few can be seen in recent
near-infrared data.  For the three brightest mid-infrared sources, we
also present 18.75 $\micron$ photometry obtained with the Keck
telescope.  The most prominent blueshifted outflows in the Orion
nebula at visual wavelengths such as HH~202, HH~203/204, HH~529, and
HH~269 all originate from OMC1-S.  The brightest infrared source in
OMC1-S at 11.7 $\micron$ is located at the base of the prominent jet
that powers HH~202 and is likely to be the sought-after driver of this
outflow. The second brightest infrared source is located at the base
of the HH~529 jet.  We consider the possibility that HH~203/204 and
HH~269 trace parts of a single bent outflow from the third-brightest
infrared source.  While there may be some lingering ambiguity about
which infrared stars drive specific jets, there is now a sufficient
number of embedded sources to plausibly account for the multiple
outflows from OMC1-S.

\end{abstract}


\section{INTRODUCTION}

Two luminous star-forming cloud cores lurk immediately behind the main
ionization front of the Orion nebula and the Trapezium cluster: the
$\sim$10$^5$ L$_{\odot}$ BN/KL nebula in the OMC1 cloud core and the
$\sim$10$^4$ L$_{\odot}$ OMC1 South region, both of which are hidden
at visual wavelengths.  The BN/KL nebula is a bright and complex
region associated with a spectacular wide-angle bipolar outflow,
powerful masers, and ultra-compact H~{\sc ii} regions, and has been
studied in great detail at infrared (IR) wavelengths (e.g., Allen \&
Burton 1993; Gezari et al.\ 1998; Shuping et al.\ 2004; Kaifu et
al. 2000; see O'Dell 2001 for a recent review).

Were it not upstaged by its spectacular northern neighbor, OMC1-S
would be the most luminous star-forming core and the most dramatic
outflow source in the Orion molecular cloud complex.  OMC1-S resides
within a visually dark and protruding region of the Orion nebula
$\sim$60$\arcsec$ southwest of the Trapezium (Wen \& O'Dell 1995).
First noted as an extended far-infrared source (Keene, Hillenbrand, \&
Whitcomb 1982), it harbors the luminous far-IR/sub-mm source FIR4
(Mezger et al.\ 1990), the dense molecular condensation CS3 (Mundy et
al.\ 1986), several compact water masers (Gaume et al.\ 1998), and at
least two bipolar molecular outflows with nearly orthogonal axes
(Ziurys, Wilson, \& Mauersberger 1990; Schmid-Burgk et al.\ 1990;
Rodriguez-Franco et al.\ 1999a, 1999b).  Additionally, Bally, O'Dell,
\& McCaughrean (2000) noted that at least a half-dozen highly
collimated Herbig-Haro (HH) outflows emerge from within OMC1 South,
giving rise to shocks best seen in {\it Hubble Space Telescope} ({\it
HST}) or Fabry-Perot images (see O'Dell et al. 1997; O'Dell \& Doi
2003; Rosado et al.\ 2001).  Proper motions of these HH objects
determined by {\it HST} indicate dynamical ages of several hundred to
a few thousand years (Bally et al.\ 2000; O'Dell \& Doi 2003).  Gaume
et al.\ (1998) discovered near-IR point sources within
$\sim$15$\arcsec$ of FIR4, offering a possible source for the embedded
CO jet discovered by Schmid-Burgk et al.\ (1990).  However, no
compact IR sources have yet been reported as plausible drivers for
most of the numerous optically-visible HH jets that emerge from
OMC1-S.

In this Letter we report the discovery of several thermal-IR point
sources in the OMC1-S cloud core.  We postulate that some of
these drive optical HH outflows from the region, since they are
located at the expected origin points of the jets deduced from
proper-motions and from new Fabry-Perot images.

\section{OBSERVATIONS}

\subsection{Thermal-IR Images}

Images of the Orion nebula at 8.8 $\micron$ ($\Delta\lambda$=8.35-9.13
$\micron$) and 11.7 $\micron$ ($\Delta\lambda$=11.09-12.22 $\micron$)
were obtained on 2004 Jan 26 and 25, respectively, using T-ReCS on
Gemini South.  T-ReCS is the facility mid-IR imager and spectrograph
with a 320$\times$240 pixel Si:As IBC array, a pixel scale on the 8m
Gemini South telescope of 0$\farcs$089, and a resulting field-of-view
of 28$\farcs$5$\times$21$\farcs$4.  The observations were taken with a
15$\arcsec$ east-west chop throw.  Since the throw is much smaller
than the extent of the nebula, we took a series of east-west scans
while chopping, starting on relatively blank sky to the west, and
stepping by 15$\arcsec$ per pointing.  To define the reference sky to
subtract from each position, we took the minimum of two frames on
either side of the position of interest.  This allowed effective
removal of point sources in the reference sky frames, but some nebular
emission persisted where both adjacent pointings contained diffuse
emission at the same position on the array. This technique worked
sufficiently well in the OMC1-South region, but degraded the image
quality somewhat because of small pointing errors in the off-source
beam.  Individual sky-subtracted frames were combined to make a larger
mosaic image, using point sources in each frame for spatial alignment.

Figure 1$a$ shows a portion of the resulting 11.7~$\micron$ mosaic
image for the region of the OMC1-South molecular cloud core southwest
of the Trapezium.  Table 1 lists J2000 coordinates for IR sources in
this field, numbered 1--11, as well as point-source photometry at 8.8
and 11.7 $\micron$ measured in a 0$\farcs$9 radius synthetic aperture.
These coordinates were measured with a 2-D radial profile fit relative
to $\theta^1$Ori C (included in the surveyed region but not shown in
Fig.\ 1$a$), assuming a position for $\theta^1$Ori C of
$\alpha_{2000}$=5$^{\rm h}$35$^{\rm m}$16$\fs$46,
$\delta_{2000}$=$-$5$\arcdeg$23$\arcmin$23$\farcs$0 (McCaughrean \&
Stauffer 1994).  The relative positional uncertainty is better than 1
pixel ($\sim$0$\farcs$09).  Flux densities in Table 1 were measured
with respect to the secondary standard star HD~32887, adopting the
values tabulated by Cohen et al.\ (1999).  Point-source sensitivity
was approximately 8.3 and 10.5 mJy (1$\sigma$) at 8.8 and 11.7
$\micron$, respectively, although uncertainty for the brighter sources
is dominated by $\pm$5--10\% uncertainty in the calibration stars.

In addition, we have obtained images of the OMC1-S region at 18.75
\micron\ ($\Delta\lambda = 18.3 -19.2$) using the Long Wavelength
Spectrometer (LWS) at the Keck Observatory on 2002 Nov 16.  LWS
employs a 128$\times$128 As:Si BIB array with a pixel scale of
0$\farcs$081, yielding a 10$\farcs$2$\times$10$\farcs$2 field of view
(Jones \& Puetter 1993).  Weather conditions were moderately stable
during the second half of the night, when OMC1-S was imaged.  We
employed a standard mid-IR chop-nod technique, with the chopping
secondary driven at 2 and 5 Hz with 20\arcsec\ and 10\arcsec\ E--W
throws, respectively.  The standard star $\beta$~And was also observed
throughout the night for PSF determination and flux calibration.  Flux
densities for sources 1--3 at 18.75~\micron\ are listed in Table~1.
Sources 1 and 2 were measured with a 15-pixel (1$\farcs$21) radius
aperture; while Source 3 was measured using an 8-pixel (0$\farcs$648)
aperture to reduce the contribution of faint diffuse emission.  Each
source was observed in several different frames and the frame-to-frame
variation (which we attribute to weather) was $\sim$5\%.  The total
uncertainty for sources 1 and 3 is $\la$7\%, while for source 2 it is
16\% due to the uncertain local background level.  Because of local
diffuse emission surrounding source 2, the reported flux density is
aperture dependent.

Sources 7-9 and 11 have visual-wavelength counterparts, appearing
stellar or unresolved in {\it HST} images (O'Dell \& Wong 1996).  The
fact that these ``naked'' stars have detectable thermal-IR emission
suggests that they have retained dusty disks not resolved by {\it
HST}.  Sources 1-6 and 10 are invisible at optical wavelengths and are
most likely embedded protostellar sources, although the position of
source 10 lies outside the OMC1-S cloud core.  This is the first
reported measurement of these sources at thermal-IR wavelengths.  Of
these, only sources 4 and 5 have been reported previously in the
near-IR (sources B and C; Gaume et al.\
1998).\footnotemark\footnotetext{Source TCC009 of McCaughrean \&
Stauffer (1994) is within 1$\arcsec$ of our source 3, but it is not
clear that these are the same source because more recent data give the
impression that the near-IR source may be a reflection nebula.  A
close inspection of the near-IR Subaru image of Orion by Kaifu et
al. (2000) reveals that all mid-IR sources except 3 and 6 have
perceptible 2 $\micron$ counterparts, although these were unremarked
by Kaifu et al.}

Thermal-IR spectral energy distributions (SEDs) for sources 1-6 are
shown in Figure 1$b$.  Blackbody curves are shown for a rough
comparison.  It is clear that sources 1 and 2 are more luminous and
more deeply embedded than sources B and C.  Sources 1, 2, 3, and 6 are
good candidates for deeply embedded outflow sources.  Of course, the
data shown in Figure 1 provide underestimates of the true bolometric
luminosities, since dust having a broad range of temperatures may emit
at wavelengths not observed here.  Furthermore, the far-IR luminosity
of the region ($\sim$10$^4$ L$_{\odot}$; Johnstone \& Bally 1999) is
much higher than the sum of the mid-IR luminosities inferred from our
data.  Also note that the blackbody curves in Figure 1 and the color
temperatures and luminosities in Table 1 do not account for silicate
absorption or reddening.

\subsection{Visual Wavelength Fabry-Perot Images}

In order to determine whether our IR sources are associated with HH
flows, we compare our T-ReCS data with Fabry-Perot images obtained by
Bally et al.\ (2004) using the Multi-Object Spectrograph/Fabry-Perot
at the f/8 Cassegrain focus of the 3.6-meter Canada-France-Hawaii
Telescope (CFHT).  For details of the observations and an analysis of
the results, see Bally et al. (2001, 2004).  The rectified data cube
has a velocity resolution of about 13 km s$^{-1}$, a pixel scale of
0$\farcs$44 pixel$^{-1}$, and a free spectral range of 392 km
s$^{-1}$.

Figure 2 shows an image extracted from a portion of the [O~{\sc iii}]
$\lambda$5007 data cube of the central Orion nebula, color-coded with
velocity, such that features near the systemic velocity are red,
low-velocity blueshifted features are green, and fast blueshifted
features are blue.  Figure 2 also illustrates the positions of the
embedded IR sources 1--6 in OMC1-S from Figure 1 and Table 1.

\section{OUTFLOWS AND THEIR SOURCES}

We have detected several closely-spaced thermal-IR sources in OMC1
South, while Bally et al.\ (2000) noted at least six optical HH jets
that emerge from sources hidden within the cloud core (HH~202, HH~529,
HH~203/204, HH~528, HH~530, and HH~269).  HH~625, an additional flow
found by O'Dell \& Doi (2003), and at least two molecular outflows,
also originate in OMC1-S.  Here we discuss possible associations
between these various outflow systems and the embedded IR sources that
we have detected.

{\it HH 202}: This is the most prominent HH object in Figure 2, and it
was one of the first in the Orion nebula to be recognized as such
(also named M42-HH2; Cant\'{o}\ et al.\ 1980; Meaburn 1986).  FP data
(Fig.\ 2; O'Dell et al.\ 1997; Rosado et al.\ 2001; Bally et al.\
2004) show a highly collimated jet emerging from OMC1-S and extending
to the brightest part of HH~202.  Source 1 lies almost exactly along
the presumed axis of the HH~202 jet within 5$\arcsec$ of where the
irradiated jet body emerges from the cloud, and it is the best
candidate for the driver of the HH202 jet.  If so, then proper motions
(O'Dell \& Doi 2003) would indicate a dynamical age of 2300$\pm$500
yr.  While we cannot positively rule out the possibility that either
source 2 or 3 drives the HH202 jet, we note that source 2 probably
drives HH529 and source 3 may drive both HH203/204 and HH269 (see
below).  To assign either source 2 or 3 the additional burden of
driving HH202 (the most prominent jet in the Orion nebula) seems less
straightforward than attributing it to the most luminous IR source in
OMC1-South (source 1), which is not a plausible driving source for any
other blueshifted HH jet.

{\it HH 529}: The HH~529 jet extends east from an embedded source in
OMC1-S at position angle PA $\approx$ 100\arcdeg .  As shown in Figure
2, IR source 2 is located exactly along the jet axis a few arcseconds
west of where the optical jet originates.  No other IR source lies
along the jet axis, so source 2 is the best candidate for the origin
point of HH529.  O'Dell \& Doi (2003) deduced the expected location of
the driving source responsible for HH~529 from proper motions to be at
5$^{\rm h}$35$^{\rm m}$14$\fs$56, --5$\arcdeg$23$\arcmin$54$\arcsec$,
where no optical source is seen.  This location is within 3$\arcsec$
of source 2.  If source 2 is the origin of HH~529, then the jet has a
dynamical age of 1100$\pm$150 yr.  O'Dell \& Doi also suggested that
this ``optical outflow source'' (OOS) is responsible for driving the
HH~269 jet (the presumed counterjet to HH~529), and perhaps HH~202,
HH~528, and HH~203/204.

{\it HH 203/204 and HH~269}: The HH~203/204 shocks trace a blueshifted
outflow extending southeast of OMC1-S at PA$\simeq$135$\arcdeg$.
Unlike HH~202 and 529, this jet cannot be traced all the way back to
OMC1-S; it becomes invisible about 1\arcmin\ southeast of this core.
The jet beam also seems to point somewhat east of OMC1-S.  Rosado et
al. (2001) proposed that HH~203/204 and HH~202 trace opposite lobes of
a bipolar outflow emerging form OMC1-South.  However, both lobes
exhibit large blueshifts, requiring a C-shaped or bent outflow
structure.  Although Bally et al.\ (2000, 2001) report several bent HH
flows in the Orion nebula, all exhibit bends in the plane of the sky,
consistent with deflection by a wind emerging from the nebular core.
If HH~203/204 and 269 originate from OMC1-South, deflection by a
stellar wind from the Trapezium stars or the plasma flowing way from
the nebula's main ionization front would deflect the flow toward the
southwest.

We propose that instead of being the counterlobe to HH~202, the
HH~203/204 pair is the deflected counterlobe to HH~269.  An arc
extending through these objects passes through IR source 3, located
several arc seconds south of source 2.  Therefore, it may be the case
that this object powers a bipolar outflow deflected toward the
southwest and toward our line of sight.  The morphology, radial
velocity, and proper motions of HH~203/204 support this hypothesis, as
discussed further by Bally et al. (2004).

{\it HH~528}: We have argued above that HH~203/204 is not the
counterjet to HH~202 as suggested intially by Rosado et al.\ --- but
if HH~202 does have a visual-wavelength counterjet (instead of one
that simply burrows into the background molecular cloud and
disappears), then the best candidate may be HH~528.  HH~528 has a
low-velocity redshift (Bally et al.\ 2004), and may be lost in the
bright emission from the ionization front if it is moving nearly in
the plane of the sky. The position angle of HH~528 is
140$\arcdeg$--150$\arcdeg$ with respect to source 1, 2, or 3, which
makes it a plausible companion to the HH~202 jet at
P.A.$\simeq$320$\arcdeg$.  However, the dynamical age of $\sim$5000 yr
(Bally et al.\ 2000) still makes it somewhat problematic to link
HH~528 and the younger HH~202 jet as part of the same bipolar flow,
unless HH~528 has been decelerated.  Alternatively, HH~528 could
potentially be associated with the fast molecular outflow source and
HH~625, discussed below.

{\it Embedded molecular outflows, HH~530, and HH~625}: Two different
molecular outflows have been detected in OMC1-S; one fast and one
relatively slow, with orthogonal axes.  The fast ($\sim$110 km
s$^{-1}$) bipolar CO jet is oriented southeast/northwest
(Rodriguez-Franco et al.\ 1999b).  Rodriguez-Franco et al.'s expected
position for this fast molecular outflow source (labeled ``FMOS'' in
Fig.\ 1) is coincident with the near-IR source ``A'' (Gaume et al.\
1998).  We detect no thermal-IR source at this position, but given the
spatial resolution of the original data, source B could easily be the
culprit as well.  In fact, in an earlier analysis of the same data,
Rodriguez-Franco et al.\ (1999a) gave a different point of origin for
the FMOS located several arcseconds to the southeast.  O'Dell \& Doi
(2003) suggested that HH~625 is where the blueshifted component of
this flow breaks out into the H~{\sc ii} region.  Since OMC1-S
protrudes from the main ionization front of the Orion nebula (Wen \&
O'Dell 1995), it is conceivable that HH~528 could be where the
redshifted part of this same bipolar flow escapes from the molecular
cloud.

The second embedded molecular outflow is a highly-collimated,
low-velocity redshifted flow toward the southwest from OMC1-S,
with a poorly-collimated blueshifted counterflow (Schmid-Burgk et
al.\ 1990). The origin of this jet is usually attributed to FIR4 or
CS3 (Mezger et al.\ 1990; Mundy 1986).  Bally et al.\ (2000) proposed
that HH~530 (see Fig.\ 2) is an optical counterpart to the
low-velocity CO jet, and proper motions indicate an origin near
FIR4+CS3 as well.  Since one of the most important outflow sources in
OMC1-S is apparently located in the vicinity of FIR4+CS3, it is
interesting that we detect no thermal-IR source there at 8.8 or 11.7
$\micron$.  This driving source must be more deeply embedded than the
others in OMC1-S, and higher-resolution observations at longer
IR, submm, and radio wavelengths are needed to measure its position
and constrain its physical properties.

In summary, we have discovered several embedded thermal-IR sources in
OMC1-S, some of which we identify as likely driving sources for HH
jets seen at visual wavelengths.  This overal picture of several
individual sources driving independent highly-collimated outflows with
different dynamical ages is distinct from the high-luminosity BN/KL
outflow 90$\arcsec$ to the north, which is a wide angle outflow that
probably originated from a single explosive event, forming multiple HH
objects at the tips of the outflowing fingers.

\acknowledgements  \scriptsize

Support for N.S.\ was provided by NASA through grant HF-01166.01A
from the Space Telescope Science Institute, which is operated by the
Association of Universities for Research in Astronomy, Inc., under
NASA contract NAS~5-26555.  Additional support was provided by NSF
grant AST 98-19820 and NASA grants NCC2-1052 and NAG-12279 to the
University of Colorado.


\begin{deluxetable}{llccccccc}
\footnotesize
\tighten
\tablewidth{0pt}
\tablenum{1}
\tablecaption{IR Point Sources in OMC1-S}
\tablehead{
 \colhead{IRS} &\colhead{Name} &\colhead{R.A.} &\colhead{DEC} &\colhead{8.8 $\mu$m}
  &\colhead{11.7 $\mu$m} &\colhead{18.8 $\mu$m} &\colhead{T$_{C}$} &\colhead{L$_{IR}$} \\
 \colhead{\ } &\colhead{\ } &\colhead{(J2000)} &\colhead{(J2000)} &\colhead{F$_{\nu}$(Jy)} 
  &\colhead{F$_{\nu}$(Jy)} &\colhead{F$_{\nu}$(Jy)} &\colhead{(K)} &\colhead{(L$_{\odot}$)}
}
\startdata
1	&138-340	&5:35:13.80  &-5:23:40.3 &3.6	    &6.4	&7.8	&290	  &13.6	 \\
2	&144-351	&5:35:14.40  &-5:23:51.0 &0.75	    &2.5	&5.8	&205	  &7.9	 \\
3	&145-356	&5:35:14.54  &-5:23:56.2 &0.21	    &0.46	&1.5	&200	  &1.9	 \\
4	&B		&5:35:13.58  &-5:23:55.5 &0.22	    &0.13	&\nodata	&980	  &1.5	 \\
5	&C		&5:35:13.55  &-5:23:59.9 &0.077	    &0.082	&\nodata	&460 &(0.2)\tablenotemark{a} \\
6	&139-357	&5:35:13.88  &-5:23:57.4 &0.017	    &0.040	&\nodata	&230	  &0.1	 \\
7	&126-344	&5:35:12.60  &-5:23:44.4 &0.21	    &0.22	&\nodata	&\nodata  &\nodata \\
8	&122-348	&5:35:12.28  &-5:23:48.4 &0.10	    &0.11	&\nodata	&\nodata  &\nodata \\
9	&134-340	&5:35:13.44  &-5:23:40.4 &0.070	    &0.096	&\nodata	&\nodata  &\nodata \\
10	&134-330	&5:35:13.40  &-5:23:29.5 &0.18	    &0.19	&\nodata	&\nodata  &\nodata \\
11	&144-334	&5:35:14.39  &-5:23:33.8 &0.23	    &0.28	&\nodata	&\nodata  &\nodata \\

\enddata \tablenotetext{a}{This mid-IR luminosity is an underestimate
for the total luminosity of C, as near-IR data (Gaume et al.\ 1998)
combined with our data would indicate a value closer to 1
L$_{\odot}$.}
\end{deluxetable}

\begin{figure}
\epsscale{0.99}
\plotone{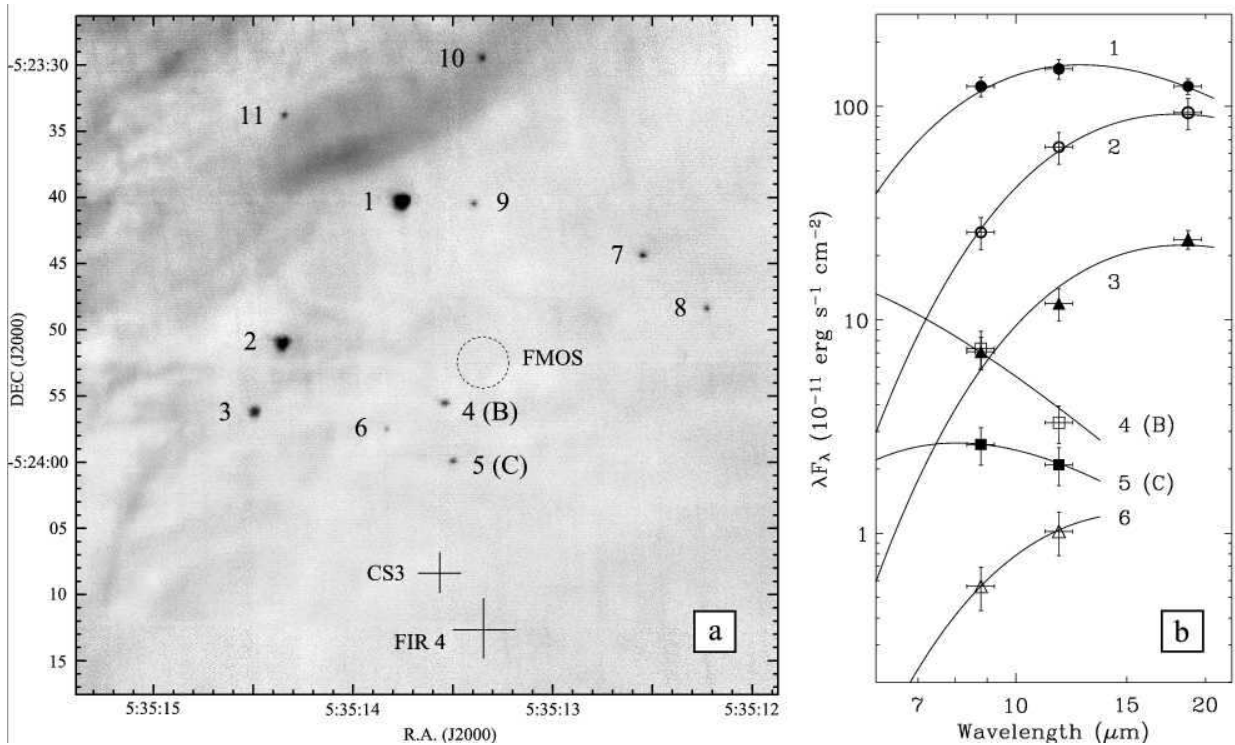}
\caption{(a) 11.7 $\micron$ image mosaic of the OMC1-S region.
Sources 1-11 are listed in Table 1.  Reported positions of FIR4
(Mezger et al.\ 1990), CS3 (Mundy et al.\ 1986), and the approximate
position of the fast molecular outflow source (FMOS; Rodriguez-Franco
et al.\ 1999b) are also indicated.  The position of source A (Gaume et
al.\ 1998) is located within the FMOS circle.  (b) Spectral energy
distributions for embedded sources in OMC1-S.  Fluxes are taken
from Table 1.  Representative Planck functions are drawn through the
data points, with corresponding temperatures and luminosities given in
Table 1.}
\end{figure}

\begin{figure}
\epsscale{0.99}
\plotone{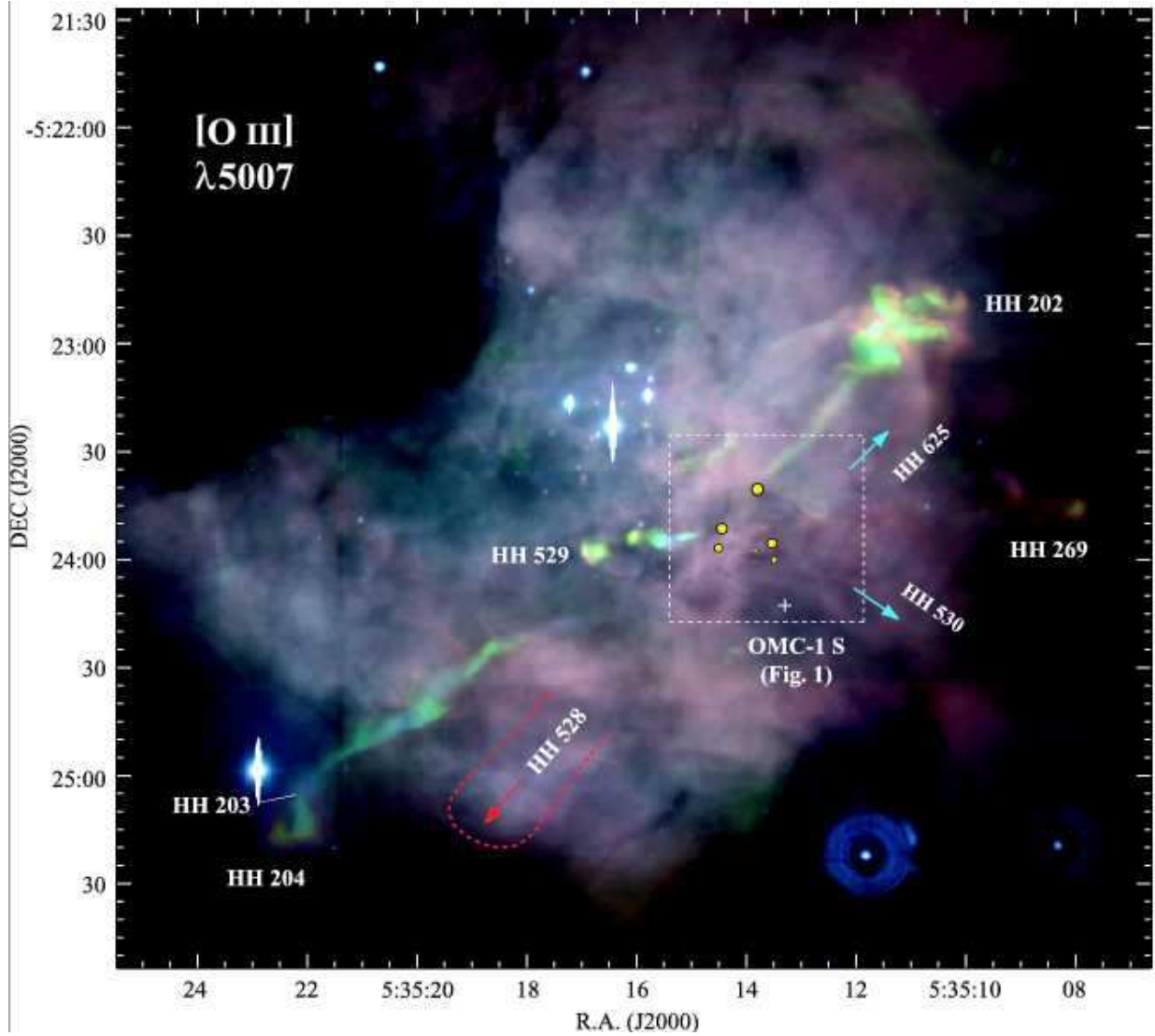}
\caption{Fabry-Perot image of the Orion nebula in [O~{\sc iii}]
$\lambda$5007 (see Bally et al.\ 2004), color-coded by velocity so
that features near the systemic velocity appear red, low-velocity
blueshifts appear green, and fast blueshifted features are blue.
Positions of embedded IR sources 1--6 from Fig.\ 1 in OMC1-S are
marked with yellow dots, and major HH jets are labeled.}
\end{figure}

\end{document}